\documentclass[%
 reprint,
superscriptaddress,
 amsmath,amssymb,
 aps,
pra,
longbibliography,
]{revtex4-2} 

\usepackage{graphicx}
\usepackage{dcolumn}
\usepackage{bm}
\usepackage{xcolor}
\usepackage{mathtools}

\begin{document}


\title{Dephasingless laser wakefield acceleration in a plasma waveguide}

\author{J.P. Palastro}
\email{jpal@lle.rochester.edu}
\affiliation{
University of Rochester, Laboratory for Laser Energetics, Rochester, New York 14623-1299 USA}

\author{K.G. Miller}
\email{kmill@lle.rochester.edu}
\affiliation{
University of Rochester, Laboratory for Laser Energetics, Rochester, New York 14623-1299 USA}

\author{C.D. Arrowsmith}
\affiliation{
University of Rochester, Laboratory for Laser Energetics, Rochester, New York 14623-1299 USA}

\author{R. Almeida}
\affiliation{GoLP/Instituto de Plasmas e Fusão Nuclear, Instituto Superior Técnico, Universidade de Lisboa, Lisbon 1049-001, Portugal}

\author{M. R. Edwards}
\affiliation{Department of Mechanical Engineering, Stanford University, Stanford, California 94305, USA}

\author{A.L. Elliott}
\affiliation{
University of Rochester, Laboratory for Laser Energetics, Rochester, New York 14623-1299 USA}

\author{A. Kiewel}
\affiliation{
University of Rochester, Laboratory for Laser Energetics, Rochester, New York 14623-1299 USA}

\author{A. Konzel}
\affiliation{
University of Rochester, Laboratory for Laser Energetics, Rochester, New York 14623-1299 USA}

\author{L.S. Mack}
\affiliation{
University of Rochester, Laboratory for Laser Energetics, Rochester, New York 14623-1299 USA}

\author{D. Ramsey}
\affiliation{
University of Rochester, Laboratory for Laser Energetics, Rochester, New York 14623-1299 USA}

\author{D. Singh}
\affiliation{Department of Mechanical Engineering, Stanford University, Stanford, California 94305, USA}

\author{A.G.R. Thomas}
\affiliation{
G\'{e}rard Mourou Center for Ultrafast Optical Science, University of Michigan, Ann Arbor, Michigan 48109, USA}

\author{J. Vieira}
\affiliation{GoLP/Instituto de Plasmas e Fusão Nuclear, Instituto Superior Técnico, Universidade de Lisboa, Lisbon 1049-001, Portugal}

\date{\today}

\begin{abstract}
Laser wakefield accelerators (LWFAs) provide extremely large accelerating gradients for compact electron accelerators and photon sources but are limited by dephasing, where trapped electrons outrun the accelerating phase of the wakefield. Flying-focus pulses can eliminate dephasing by driving a wake at the vacuum speed of light, but these pulses involve tradeoffs such as varying spot size, long duration, or large plasma volume. Here we show that a spatiotemporally structured laser pulse propagating in a plasma waveguide can drive a wakefield at the vacuum speed of light while maintaining a constant spot size and ultrashort duration. The pulse is formed by superposing plasma-waveguide modes with appropriately selected frequencies. Compared with flying-focus approaches, the waveguide substantially reduces the required plasma volume. Scaling laws and quasi-3D particle-in-cell simulations show that the single-stage energy gain increases linearly with the number of modes used to construct the pulse, enabling larger energy gains or shorter stages than standard LWFA. 
\end{abstract}
              
\maketitle

\section{Introduction}
In laser wakefield acceleration (LWFA), a high-intensity laser pulse drives a large-amplitude plasma wave that traps and accelerates electrons to high energies \cite{Tajima1979,Faure2004laser, Geddes2004,Mangles2004,Lu2007,Esarey2009,Miao2022,Picksley2024}. The longitudinal electric field of the plasma wave exceeds that of conventional radio-frequency accelerators by nearly three orders of magnitude, positioning LWFA as a potential technology for compact particle colliders and light sources. Despite this potential, LWFA faces limitations, such as dephasing, in which trapped electrons outrun the accelerating phase of the wakefield. This arises from the mismatch between the subluminal group velocity of the driving laser pulse, which determines the phase velocity of the wake, and the near-luminal velocity of the high-energy electrons. Dephasing can be overcome by accelerating electrons through a series of LWFA stages \cite{Steinke2016,Luo2018,Pathak2018,Thomas2021,Schroeder2023}, but this introduces significant technological challenges, including timing, alignment, and preservation of electron-beam quality between stages. These challenges motivate the development of concepts that mitigate dephasing to reduce the number of stages or eliminate the need for staging altogether \cite{Sprangle2001,Yoon2014,Debus2019,Palastro2020,Caizergues2020,Palastro2021,Miller2023,Pierce2025,Arrowsmith2025,Liberman2026}.

Flying-focus pulses \cite{Sainte-Marie2017,Froula2018,Jolly2020,Palastro2020,Ambat2023,Pigeon2023,Li2024,Piccardo2025,Cao2026} circumvent dephasing by driving a wake at the vacuum speed of light, ensuring that high-energy electrons remain in the accelerating phase of the wakefield \cite{Palastro2020,Caizergues2020,Miller2023,Pierce2025,Arrowsmith2025,Liberman2026}. These pulses feature an intensity peak that travels at a programmable velocity over distances far greater than a Rayleigh range while maintaining a near-constant profile. The ``dephasingless" laser wakefield acceleration (DLWFA) enabled by flying-focus pulses allows for either larger energy gains in a stage of fixed length or equal energy gains in a shorter stage when compared to standard LWFA. Optical techniques for generating flying-focus pulses control the focal time and location of each frequency, time slice, sub-pulse, or annulus of the overall pulse. However, each technique involves tradeoffs in the resulting pulse structure, such as a longitudinally varying spot size, extended pulse duration, or constraints on the transverse or temporal profiles. 

The ultrashort flying focus \cite{Palastro2020,Caizergues2020,Pigeon2023}, for instance, uses an axiparabola to focus different annuli to different longitudinal locations \cite{Smartsev2019,Geng2022,Oubrerie2022} and a radial echelon \cite{Palastro2020,Ambat2023,Pigeon2023} to control the relative timing of the annuli. While the geometric aberration of the axiparabola has the benefit of extending the focal range, it also produces a longitudinally varying spot size \cite{Pigeon2023,Liberman2025}. Moreover, achieving the ultrashort intensity peaks optimal for LWFA necessitates an axiparabola with a relatively small, full-aperture $f$-number (e.g., $f_\# \sim 5$). As a result, the volume of plasma $V$ needed to avoid plasma refraction and achieve the desired focal velocity increases rapidly with the accelerator length $L$, scaling as $V \propto L^3/f_\#^2$. This is a critical challenge to realizing ${\sim}100$-GeV energy gains in a single, $L \sim 1$-m stage \cite{Shaw2025}. Nonetheless, flying-focus--driven DLWFA has been experimentally demonstrated at shorter acceleration lengths, $L \sim 1 \;\mathrm{cm}$, and shows promise for scaling to $L \sim 10 \;\mathrm{cm}$ \cite{Arrowsmith2025}.

\begin{figure*}
\includegraphics[width=0.9\textwidth]{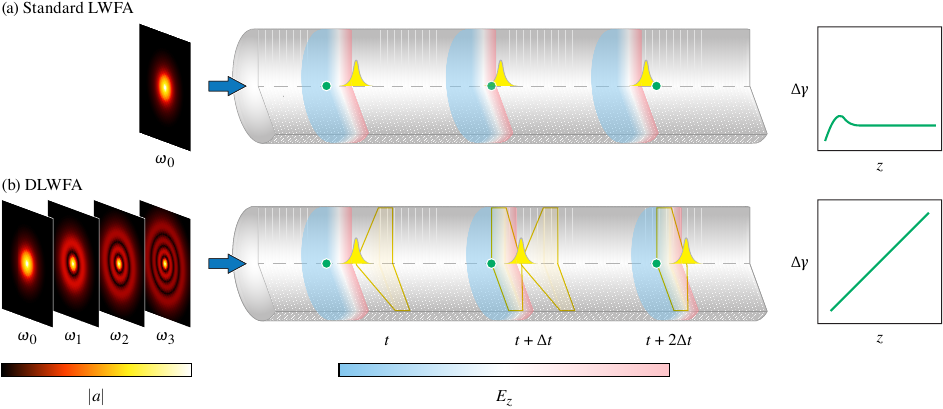}
\caption{Schematic of standard and dephasingless laser wakefield acceleration in a plasma waveguide. (a) In standard LWFA, an ultrashort laser pulse (yellow) with a single characteristic frequency $\omega_0$ and Gaussian transverse profile (red scale) enters a plasma waveguide (gray) and drives a longitudinal wakefield $E_z$. High-energy electrons (green), initially located in the accelerating phase of the wake (blue), gain energy until they advance into a decelerating phase (red), where they lose energy (right). Eventually, the electrons outrun the wake and pulse altogether. (b) In DLWFA, a laser pulse composed of waveguide modes with distinct frequencies enters the waveguide, where modal interference produces an ultrashort intensity peak that drives a longitudinal wakefield at the vacuum speed of light. High-energy electrons remain in the accelerating phase of the wake and gain energy over the entire length of the accelerator. The yellow structure depicts the overall envelope of the DLWFA pulse, which travels at the subluminal group velocity.}
\label{fig:f1}
\end{figure*}

Here, we introduce DLWFA driven by a spatiotemporally structured laser pulse in a plasma waveguide (Fig. \ref{fig:f1}). The approach combines the velocity control of the flying focus with the established benefits of plasma waveguides \cite{Durfee1993,Ehrlich1996,Ditmire1998,Clark2000,Kumarappan2005,Geddes2005,Layer2007,Shalloo2018,Miao2020,Palastro2025} to reduce the required plasma volume. The pulse is constructed from plasma-waveguide modes with appropriately selected frequencies, producing an intensity peak that travels at the vacuum speed of light while maintaining a constant spot size and ultrashort duration. Owing to the narrow transverse profile of the plasma waveguide, the required plasma volume can be substantially smaller than in previous flying-focus approaches. The electron energy gain and stage length increase linearly with the number of waveguide modes used to construct the pulse, enabling either larger energy gains per stage or shorter stages relative to standard LWFA. In the single-mode limit, the scalings reduce to those of standard LWFA, whereas the two-mode limit resembles beat-wave--driven LWFA \cite{Everett1994,Pukhov2023}. For a pulse composed of 10 modes, quasi-3D particle-in-cell (PIC) simulations demonstrate acceleration through a stage $11\times$ longer than the dephasing length, yielding a corresponding $11\times$ enhancement in the electron energy gain compared to standard LWFA at the same density.

The remainder of this article is organized as follows. Section II reviews the theory of arbitrary-velocity laser pulses in plasma waveguides \cite{Palastro2025}. Section III describes the design of a pulse and plasma waveguide for DLWFA. Section IV presents scaling laws and compares DLWFA in a plasma waveguide with standard LWFA, highlighting the advantages and tradeoffs. Section V demonstrates DLWFA with quasi-3D PIC simulations and validates the scaling laws. Section VI concludes the article with a summary of the results and a discussion of future prospects.

\section{Arbitrary-Velocity Laser Pulses in Plasma Waveguides}

Consider a laser pulse propagating in the positive $\hat{\mathbf{z}}$ direction through a preformed, parabolic plasma waveguide. The electron density of the waveguide $n(r)$ can be expressed in terms of the squared plasma frequency $\omega_\mathrm{p}^2(r) = e^2 n(r)/m\varepsilon_0$ and parameterized as
\begin{equation}\label{eq:profile}
\omega_\mathrm{p}^2(r) = \omega_{\mathrm{p}0}^2 + \frac{4c^2}{w^4}r^2, 
\end{equation}
where $r=(x^2+y^2)^{1/2}$ is the radial distance from the $z$ axis, $\omega_\mathrm{p0}^2 \equiv e^2 n(0)/m\varepsilon_0$, and $w$ is the ``matched'' spot size. The normalized, transverse vector potential of the pulse, $\mathrm{a} = eA/mc$, can be decomposed into positive- and negative-frequency components $\mathrm{a} = \tfrac{1}{2}(a+a^*)$, where
\begin{equation}\label{eq:FT}
a(\mathbf{x},t) = \frac{1}{2\pi}\int \mathrm{e}^{-i\omega t}\tilde{a}(\mathbf{x},\omega) d\omega,
\end{equation}
and the integral is taken over positive frequencies. Each spectral component is a superposition of $N$ waveguide modes:
\begin{equation}\label{eq:sols}
\tilde{a}(\mathbf{x},\omega) = \frac{1}{N}\sum_{q} \tilde{\alpha}_{q}(\omega)L_{q}\left(\frac{2r^2}{w^2}\right)\mathrm{exp}\bigg({-\frac{r^2}{w^2}}\bigg)\mathrm{e}^{ik_{q}(\omega)z},
\end{equation}
where $q$ is the radial mode number, $L_q$ is a Laguerre polynomial, and
\begin{equation}\label{eq:kq}
k_{q}(\omega) = \frac{1}{c}\left(\omega^2 -\omega_\mathrm{p0}^2 - \frac{4(1 + 2q )c^2}{w^2}\right)^{1/2}. 
\end{equation}
The coefficients
\begin{equation}\label{eq:alpha}
\tilde{\alpha}_{q}(\omega) = |\tilde{\alpha}_{q}(\omega)|\mathrm{e}^{i\Phi_{q}(\omega)}
\end{equation}
encode the spectral amplitude $|\tilde{\alpha}_{q}(\omega)|$ and phase $\Phi_{q}(\omega)$ of each mode. 

A laser pulse with an intensity peak that travels at a prescribed velocity is constructed by appropriately tailoring the spectrum of each radial mode \cite{Palastro2025}. The appropriate spectra can be derived by analyzing the on-axis vector potential $a_\mathrm{oa}(z,t) \equiv a(r=0,z,t)$:
\begin{equation}\label{eq:FTOA}
a_\mathrm{oa}(z,t) \equiv \frac{1}{2\pi N}\sum_{q} \int \tilde{\alpha}_{q}(\omega)\mathrm{e}^{i(k_{q}(\omega)z-\omega t)} d\omega. 
\end{equation}
A vector potential whose amplitude travels at a longitudinal velocity $v_\mathrm{a}$ while maintaining its spatiotemporal profile satisfies
\begin{equation}\label{eq:sym}
a_\mathrm{oa}(z + v_\mathrm{a} \mathcal{T},t + \mathcal{T}) = a_\mathrm{oa}(z,t)\mathrm{e}^{i\eta \mathcal{T}},
\end{equation}
where $\mathcal{T}$ is an arbitrary time interval, $\eta$ is a constant, and the factor $e^{i\eta \mathcal{T}}$ accounts for uniform carrier-phase evolution. Imposing this relation in Eq.~\eqref{eq:FTOA} yields the constraint $\omega = k_q(\omega) v_\mathrm{a} + \eta$. Because each radial mode obeys a distinct dispersion relation [Eq.\eqref{eq:kq}], no single broadband spectrum can satisfy this constraint simultaneously for all $q$. However, the constraint can be satisfied if each radial mode has a distinct frequency given by 
\begin{equation}\label{eq:cond}
\omega_q = \omega_0 + \left(k_q(\omega_q) - \kappa_{0}\right)v_\mathrm{a},
\end{equation}
where $\eta = \omega_0 -\kappa_0v_\mathrm{a}$ has been chosen so that the frequency of the lowest-order mode $\omega_0$ serves as a reference frequency and $\kappa_0 \equiv k_0(\omega_0)$. Geometrically, the allowed $(k_q, \omega_q)$ pairs correspond to the intersections of the dispersion hyperbolae [Eq.~\eqref{eq:kq}] and the lines $\omega = k_qv_\mathrm{a} + \eta$ [Eq.~\eqref{eq:cond}]. The constraint expressed by Eq.~\eqref{eq:cond}  requires the spectral amplitudes to be sharply peaked at $\omega = \omega_q$. In the simplest case, as considered here, all modes share the same sharply peaked profile $|\tilde{\alpha}_{q}(\omega)| = \tilde{\alpha}(\omega-\omega_{q})$.

\section{Design of Laser pulse and plasma waveguide for DLWFA}

The energy gain in LWFA is determined by the longitudinal electric field of the wake $E_{z}$ and the acceleration length $L$: $\Delta \gamma \simeq e|E_z|L/mc^2$, where $\gamma$ is the Lorentz factor. Traditionally, the acceleration length of a single LWFA stage has been limited by dephasing, with electrons traveling close to the vacuum speed of light outrunning the accelerating phase of the wake after a distance
\begin{equation}\label{eq:Ld}
L_\mathrm{d} = \frac{\pi c^2}{\omega_\mathrm{p0}(c-v_\mathrm{w})}.
\end{equation}
The dephasing length $L_\mathrm{d}$ depends sensitively on the phase velocity of the wake $v_\mathrm{w}$, which is set by the velocity of the driving intensity peak. For standard laser pulses, this is approximately their subluminal group velocity, $v_\mathrm{w} \approx v_\mathrm{g} < c$. The goal here is to design a pulse with an intensity peak that travels at $v_\mathrm{a} = c$ and drives a wake with $v_\mathrm{w} = c$. In this limit, $L_\mathrm{d} \rightarrow \infty$, eliminating dephasing as an upper bound on the acceleration length.

When $v_\mathrm{a} = c$, the required frequencies of the radial modes are given by  
\begin{equation}\label{eq:omgq}
\omega_{q} = \omega_{0} + \frac{4c^2q}{(\omega_0 - c\kappa_0)w^2}. 
\end{equation}
For compatibility with the bandwidths of typical laser systems, the frequencies should satisfy $\omega_q-\omega_0 \ll \omega_0$. This condition can be rewritten in terms of the parameters of the plasma waveguide by substituting $\kappa_0 = k_0(\omega_0)$ into Eq.~\eqref{eq:omgq}. In the underdense ($\omega_0 \gg \omega_{\mathrm{p}}$) regime relevant to LWFA, the condition becomes
\begin{equation}\label{eq:cond2}
 \left(\frac{\omega_{\mathrm{p0}}w}{2c}\right)^2 \gg 1,
\end{equation}
which states that the matched spot size should be much larger than the skin depth. With this condition satisfied, it is convenient to define the characteristic frequency spacing
\begin{equation}\label{eq:BOmg}
\Omega \equiv 2\omega_0\bigg(\frac{2c}{\omega_\mathrm{p0}w}\bigg)^2.
\end{equation}
The required frequencies then simplify to
\begin{equation}\label{eq:omgqa}
\omega_{q} = \omega_0 + q\Omega. 
\end{equation}
Thus, an intensity peak with $v_\mathrm{a} = c$ can be achieved by incrementally shifting the frequency of each mode relative to the reference frequency $\omega_0$.

The spatiotemporal profile of the pulse is determined by the modal frequencies and spectral amplitudes $\tilde{\alpha}(\omega - \omega_q)$. The temporal structure is most clearly seen in the on-axis ($r=0$) profile
\begin{equation}\label{eq:STPOA}
|a_\mathrm{oa}(z,t)|^2 = \frac{1}{N^2} |\alpha(t-z/v_{\mathrm{g}})|^2\frac{\sin^2[\tfrac{1}{2}N\Omega(t-z/c)]}{\sin^2[\tfrac{1}{2}\Omega(t-z/c)]},
\end{equation}
where $v_\mathrm{g} = c^2k_0(\omega_0)/\omega_0$ [see Appendix A for derivation]. This expression separates the temporal envelope $\alpha(t-z/v_\mathrm{g})$ from the interference structure arising from the superposition of radial modes. The envelope propagates at the group velocity $v_\mathrm{g}$, while the intensity peaks formed by the interference travel at $c$. The peaks repeat with a period $T_\mathrm{R} = 2\pi/\Omega$, or equivalently,
\begin{equation}\label{eq:TR}
T_\mathrm{R} = \frac{ \pi }{\omega_0} \bigg(\frac{\omega_\mathrm{p0}w}{2c}\bigg)^2,
\end{equation}
and have a full-width at half-maximum (FWHM) duration 
\begin{equation}\label{eq:fwhm}
\tau = \frac{\sigma \pi }{N\omega_0} \bigg(\frac{\omega_\mathrm{p0}w}{2c}\bigg)^2,
\end{equation}
where $\sigma$ is a numerical factor that monotonically decreases from $1$ to $0.88$ as $N$ becomes large. 
Due to the velocity mismatch between the peaks and the envelope, an individual peak traverses the envelope after a distance
\begin{equation}\label{eq:L}
L_\mathrm{D} =  \bigg( \frac{cv_\mathrm{g}}{c-v_\mathrm{g}} \bigg)T \approx 2\bigg(\frac{\omega_0}{\omega_\mathrm{p0}}\bigg)^2cT, 
\end{equation}
where $T$ is the duration of the envelope.

A design for DLWFA driven by a luminal intensity peak arising from modal interference is guided by two considerations. First, the FWHM duration of the peak should be chosen to maximize the strength of the wakefield. In the linear regime ($a_0\equiv\max|a|\lesssim0.5$), this occurs for $\tau = \pi/\omega_\mathrm{p0}$. Second, the recurrence period should satisfy $T_\mathrm{R} \simeq T$ to maximize the distance that the intensity peak drives the wake before a preceding peak disrupts the downstream plasma. The first consideration determines the matched spot size through Eq.~\eqref{eq:fwhm}:
\begin{equation}\label{eq:wm}
w = \frac{2c}{\omega_\mathrm{p0}}\bigg(\frac{\omega_0}{\omega_\mathrm{p0}}\bigg)^{1/2}\bigg(\frac{N}{\sigma}\bigg)^{1/2} .
\end{equation}
While the matched spot size sets the profile of the waveguide [Eq.~\eqref{eq:profile}], the FWHM spot
\begin{equation}\label{eq:wF}
w_\mathrm{F} = \frac{2c}{\omega_\mathrm{p0}}\bigg(\frac{\omega_0}{\omega_\mathrm{p0}}\bigg)^{1/2}\bigg(\frac{2\ln(2)}{\sigma}\bigg)^{1/2}
\end{equation}
governs the width of the wake and is smaller than $w$ for $N>1$ due to the coherent superposition of radial modes. The second consideration determines the stage length through Eq.~\eqref{eq:L}:
\begin{equation}\label{eq:La}
L_\mathrm{D} =  \frac{2\pi c}{\omega_\mathrm{p0}}\bigg(\frac{\omega_0}{\omega_\mathrm{p0}}\bigg)^2\frac{N}{\sigma}. 
\end{equation}
For DLWFA in a plasma waveguide, Eq.~\eqref{eq:La} replaces the dephasing length as the upper bound on the stage length. In principle, $L_\mathrm{D}$ can be increased arbitrarily by increasing the number of modes, albeit at the expense of greater system complexity and a broader laser bandwidth, $\Delta \omega \sim (N-1)\Omega$.

\begin{figure}
\includegraphics{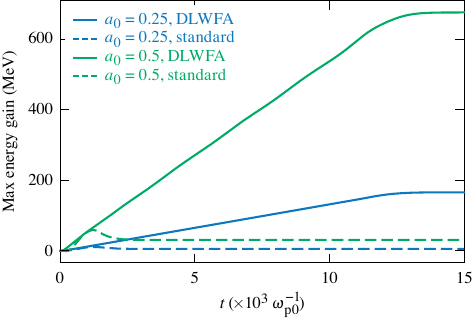}
\caption{Comparison of the maximum energy gain from DLWFA and standard LWFA in a plasma waveguide. With DLWFA (solid), an externally injected 100-MeV electron gains approximately $11\times$ the dephasing-limited energy gain, consistent with the $N=10$ modes of the pulse. With traditional LWFA (dashed), the electron reaches the dephasing-limited energy before advancing into a decelerating phase and outrunning the wake. In both cases, the maximum energy gains agree with the scalings [Eqs.~\eqref{eq:dgD} and~\eqref{eq:dgS}] for $a_0 = 0.25$ (blue) and $0.5$ (green). Simulation parameters are provided in Table I.}
\label{fig:f2}
\end{figure}

\section{Comparison with Standard LWFA}
An ultrashort intensity peak with $\tau = \pi/\omega_\mathrm{p0}$ and a mildly relativistic amplitude, $a_0\lesssim0.5$, drives a wake with a maximum longitudinal electric field 
\begin{equation}\label{eq:Ez}
\frac{e|E_z|}{mc} = \frac{\pi}{8}\omega_\mathrm{p0}a_0^2.
\end{equation}
Through $\omega_\mathrm{p0} \propto \sqrt{n_0}$, the on-axis density sets the maximum field as well as the acceleration lengths for DLWFA (D) and standard LWFA (S). To compare performance at fixed $a_0$ without requiring equal densities, the on-axis density, plasma frequency, FWHM spot size, and matched spot size are denoted $n_j$, $\omega_{\mathrm{p}j}$, $w_{\mathrm{F}j}$, and $w_j$, respectively, with $j=\mathrm{D},\mathrm{S}$. For DLWFA, the stage length is given by Eq.~\eqref{eq:La}, yielding an energy gain
\begin{equation}\label{eq:dgD}
\Delta \gamma_\mathrm{D} = \frac{\pi^2}{4}\frac{N}{\sigma}  \bigg(\frac{\omega_0}{\omega_\mathrm{pD}}\bigg)^2a_0^2.
\end{equation}
For standard LWFA, the stage length is instead limited by the dephasing length [Eq.~\eqref{eq:Ld}], $L_\mathrm{S} =L_\mathrm{d} = 2\pi c \omega_0^2 /\omega_\mathrm{pS}^3$, where $v_\mathrm{w} = v_\mathrm{g}\approx c(1-\omega_\mathrm{pS}^2/2\omega_0^2)$ has been used. The corresponding energy gain is
\begin{equation}\label{eq:dgS}
\Delta \gamma_\mathrm{S} = \frac{\pi^2}{4}  \bigg(\frac{\omega_0}{\omega_\mathrm{pS}}\bigg)^2a_0^2.
\end{equation}
In the limit of a single mode ($N=1$, $\sigma = 1$)  and equal on-axis density ($n_\mathrm{D} = n_\mathrm{S}$), the energy gains are identical: $\Delta\gamma_\mathrm{D} = \Delta\gamma_\mathrm{S}$. Thus, the advantage of DLWFA arises from the use of multiple modes and operation at a different density.


A meaningful comparison of DLWFA and standard LWFA must also consider the energy required in their respective drive pulses. Assuming a flattop profile for $\alpha$ with maximum amplitude $a_0$, the energy in the DLWFA pulse is 
\begin{equation}\label{eq:UD}
\frac{U_\mathrm{D}}{mc^2} = \frac{\pi cN}{4\sigma^2r_\mathrm{e}\omega_0} \bigg(\frac{\omega_0}{\omega_\mathrm{pD}}\bigg)^{4} a_0^2,
\end{equation}
where $r_\mathrm{e}$ is the classical electron radius.
For a standard LWFA pulse with the same normalized FWHM spot size, i.e., $\omega_{\mathrm{pD}} w_{\mathrm{F}\mathrm{D}} = \omega_{\mathrm{pS}} w_{\mathrm{F}\mathrm{S}}$, the energy is
\begin{equation}\label{eq:UT}
\frac{U_\mathrm{S}}{mc^2} = \frac{\pi c}{4\sigma r_\mathrm{e}\omega_0} \bigg(\frac{\omega_0}{\omega_\mathrm{pS}}\bigg)^{3} \bigg(\frac{\omega_0}{\omega_\mathrm{pD}}\bigg) a_0^2.
\end{equation}
As before, the pulse energies are equal in the limit of a single mode and equal on-axis density: $U_\mathrm{D} = U_\mathrm{S}$.

The relative performance of DLWFA and standard LWFA depends on whether the comparison is made at fixed density, fixed stage length, or fixed energy gain. At fixed density ($n_\mathrm{D} = n_\mathrm{S}$), the ratio of energy gains is
\begin{equation}\label{eq:fixedn}
\frac{\Delta \gamma_\mathrm{D}}{\Delta \gamma_\mathrm{S}} = \frac{N}{\sigma}.
\end{equation}
In this case, $U_\mathrm{D}/U_\mathrm{S} = N/\sigma$, reflecting the longer stage length, $L_\mathrm{D} = (N/\sigma)L_\mathrm{S}$. The matched spot sizes satisfy $w_\mathrm{D} = N^{1/2}w_\mathrm{S}$. At fixed stage length ($L_\mathrm{D} = L_\mathrm{S}$), the energy-gain ratio becomes
\begin{equation}\label{eq:fixedL}
\frac{\Delta \gamma_\mathrm{D}}{\Delta \gamma_\mathrm{S}} = \bigg(\frac{N}{\sigma} \bigg)^{1/3}.
\end{equation}
Here, $U_\mathrm{D} = U_\mathrm{S}$ and $w_\mathrm{D} = (\sigma^{2}N)^{1/6}w_\mathrm{S}$. 
Finally, at fixed energy gain ($\Delta \gamma_\mathrm{D} = \Delta \gamma_\mathrm{S}$), the ratio of stage lengths is 
\begin{equation}\label{eq:fixedG}
\frac{L_\mathrm{D}}{L_\mathrm{S}} = \bigg(\frac{\sigma}{N} \bigg)^{1/2}.
\end{equation}
In this regime, $U_\mathrm{D}/U_\mathrm{S} = (\sigma/N)^{1/2}$, which shows that the reduction in required pulse energy equals the reduction in length [Eq.~\eqref{eq:fixedG}]. The matched spot sizes satisfy $w_\mathrm{D} = \sigma^{1/2}w_\mathrm{S}$. In the latter two cases, the improved performance of DLWFA results from operation at higher on-axis density: $n_{\mathrm{D}} = (N/\sigma)^{2/3}n_{\mathrm{S}}$ for fixed stage length and $n_{\mathrm{D}} = (N/\sigma)n_{\mathrm{S}}$ for fixed energy gain.

\begin{figure*}
\includegraphics{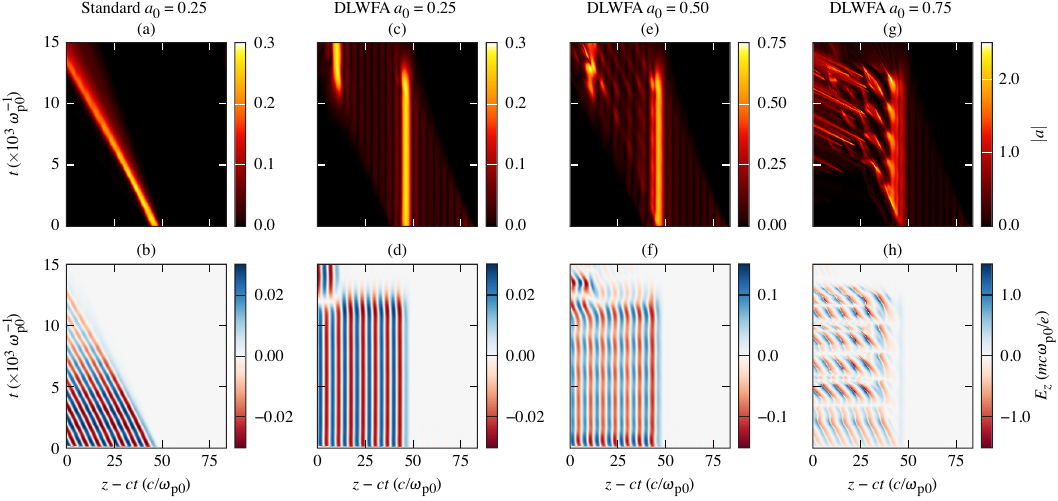}
\caption{On-axis ($r=0$) structure and evolution of the driving laser pulse and wakefield in a frame moving at the vacuum speed of light. (a) A standard pulse with $a_0 = 0.25$ drives the wake shown in (b) whose phase fronts slip backward due to the subluminal group velocity. (c-f) The luminal intensity peak of DLWFA pulses with (c) $a_0 = 0.25$ and (e) $a_0 = 0.5$ drive wakes with near-stationary phase fronts, eliminating dephasing. The first recurrence of the moving intensity peak is visible at $\omega_\mathrm{p0}t\approx10^4$. In (e), at $a_0 = 0.5$, self-focusing and depletion begin to modify the pulse profile and corresponding wake. (g) At $a_0 = 0.75$, self-focusing and depletion significantly distort the trajectory of intensity peak and the associated wake. Simulation parameters are provided in Table I.}
\label{fig:f3}
\end{figure*}

\begin{table}[b]
\caption{\label{tab:table1}
Laser pulse, plasma, and accelerator parameters for the figures and simulations. The parameters are motivated by commonly used laser systems and experimentally demonstrated plasma channels. The vacuum wavelength $\lambda_0 = 2\pi c/\omega_0$ and $\Lambda \equiv (\lambda_0/\omega_0)\Omega$. In the figures and simulations, time and space are normalized to $1/\omega_\mathrm{p0} = 7.24\,\mathrm{fs}$ and $c/\omega_\mathrm{p0} = 2.17 \,\mathrm{\mu m}$, respectively.}
\begin{ruledtabular}
\begin{tabular}{ccc}
Pulse & Standard & Dephasingless \\
\hline
$\lambda_0$ & 1 $\mu$m & 1 $\mu$m \\
$a_0$ & 0.25,\,0.5,\,0.75 & 0.25,\,0.5,\,0.75 \\
$\tau$  & 23 fs & 23 fs \\
$w_\mathrm{F}$  & $20$ $\mu$m & $20$ $\mu$m \\
$N$  & 1 & 10 \\
$\Lambda$  & - & 13 nm \\
\hline
Plasma & Standard &  Dephasingless \\
\hline
$n(0)$ & $6\times10^{18} \text{cm}^{-3}$ & $6\times10^{18} \text{cm}^{-3}$ \\
$w$ & $17$ $\mu$m & $54$ $\mu$m \\
\hline
Accelerator & Standard &  Dephasingless \\
\hline
$L_\mathrm{d}$ & 2.5 mm & - \\
$L_\mathrm{D}$ & - & $2.8$ cm \\
\end{tabular}
\end{ruledtabular}
\end{table}

\section{Simulation Results}

The analysis above describes the design of DLWFA in a plasma waveguide and the associated scaling laws. To demonstrate a representative design and confirm the scaling behavior, simulations were conducted using the particle-in-cell code \textsc{osiris} (see Table I for physical parameters and Appendix B for details). Comparisons with standard LWFA at equal on-axis density ($n_\mathrm{D} = n_\mathrm{S}$) show that $N=10$ modes result in a stage length $11\times$ the dephasing length, with an energy gain enhancement of $11\times$. The simulations also indicate that DLWFA deviates from the ideal linear regime at $a_0 \sim 0.75$ due to the onset of etching and self-focusing, which induces mode coupling and modifies the modal interference and propagation dynamics.

Figure~\ref{fig:f2} shows that the maximum energy gain of externally injected 100-MeV electrons is more than an order of magnitude larger in DLWFA (solid line) than in standard LWFA (dashed line). This enhancement is consistent with the number of modes in the DLWFA pulse and holds for both $a_0 = 0.25$ (blue) and $0.5$ (green). In DLWFA, the highest-energy electron gains energy over the entire length $L_\mathrm{D}$, whereas in standard LWFA the electron gains energy over a dephasing length before entering a decelerating phase and eventually outrunning both the wake and the pulse. In both cases, the maximum energy gains from the simulations agree closely with the scaling predictions; for example, at $a_0 = 0.5$,  $mc^2\Delta \gamma_\mathrm{D} = 660 \,\mathrm{MeV}$ and $mc^2\Delta \gamma_\mathrm{S} = 58\,\mathrm{MeV}$. 

Figure~\ref{fig:f3} compares the on-axis vector potential $|a_\mathrm{oa}|$ and longitudinal wakefield $E_z$, illustrating two key features. First, at lower amplitudes, the laser pulse and accelerating field in standard LWFA slip backward in a frame moving at the speed of light, resulting in dephasing. This contrasts with DLWFA, where the pulse amplitude and accelerating field remain nearly stationary, thereby eliminating dephasing. Second, at higher amplitudes, $a_0 \sim 0.75$, the DLWFA pulse self-focuses, depletes, redshifts, and compresses. The accompanying reduction in group velocity slows the phase velocity of the wake, culminating in a burst that dislocates the wake phase. This process repeats as downstream temporal slices of the pulse constructively interfere to form the moving intensity peak. Although nonlinear propagation in the form of temporal pulse splitting is also observed for $a_0 = 0.5$, a steady state develops, stabilizing the wake structure.

At larger values of $a_0$ (${\gtrsim0.75}$), the increased propagation length of the DLWFA pulse, due to the longer stage length, makes it more susceptible to self-focusing and depletion-induced redshifting and compression. Despite its larger matched spot size ($w_\mathrm{D} = N^{1/2}w_\mathrm{S}$), the peak power of the DLWFA pulse is identical to that of the conventional pulse. This follows directly from the modal interference producing an intensity peak with a smaller FWHM spot size [Eq.~\eqref{eq:wF}], such that $w_{\mathrm{FD}} = w_{\mathrm{FS}}$. As a consequence, the power-to-critical-power ratios are identical for the two pulses: 
\begin{equation} \label{eq:PPcr}
\bigg(\frac{P}{P_\mathrm{cr}}\bigg)_\mathrm{D} = \frac{\omega_{\mathrm{pD}}^2 w_{\mathrm{D}}^2}{32c^2N}
 a_0^2 = \frac{\omega_{\mathrm{pS}}^2 w_{\mathrm{S}}^2}{32c^2}
 a_0^2 = \bigg(\frac{P}{P_\mathrm{cr}}\bigg)_\mathrm{S}, 
\end{equation}
where $P_\mathrm{cr}$ is the critical power \cite{Sun1987}. However, because of its longer propagation distance, the DLWFA pulse accumulates more nonlinear phase and experiences stronger nonlinear refraction. In addition, as $a_0\rightarrow 1$, the depletion length  $L_\mathrm{pd} = \pi cN\omega_0^2/a_0^2\omega_\mathrm{p0}^3$ \cite{Lu2007} decreases toward the stage length $L_\mathrm{D}$. This depletion leads to redshifting, compression, and the group velocity reduction observed in Fig.~\ref{fig:f3} \cite{Zhu2013}.

\begin{figure}
\includegraphics{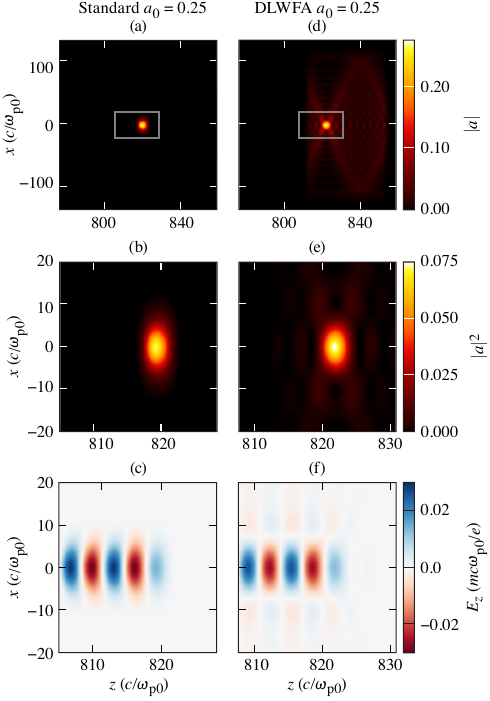}
\caption{Comparison of the spatiotemporal structure of the driving laser pulse and wakefield in standard LWFA and DLWFA at fixed density, $n_\mathrm{S} = n_\mathrm{D}$. (a,b) For a standard pulse, both the matched spot size $w_\mathrm{S}$ and FWHM spot size $w_{\mathrm{F}\mathrm{S}}$ are small, leading to a wake of comparable transverse width as shown in (c). (d,e) For a DLWFA pulse, the matched spot size is large, $w_\mathrm{D} = N^{1/2}w_\mathrm{S}$, but the FWHM spot size is small, $w_{\mathrm{F}\mathrm{D}} = w_{\mathrm{F}\mathrm{S}}$. As a result, the DLWFA pulse drives a wake with nearly the same transverse width as the standard pulse, as seen in (f). In both cases, $a_0 = 0.25$. The structure of the modal interference and its effect on the wake are apparent in (d) and (f). Simulation parameters are provided in Table I.}
\label{fig:f4}
\end{figure}

Figure~\ref{fig:f4} shows that the DLWFA pulse produces the same local features as the standard LWFA pulse despite its much larger matched spot size. Within a transverse region comparable to the FWHM spot size, both pulses have nearly identical spatiotemporal profiles and drive nearly identical longitudinal fields. Outside this region, the modal interference in the DLWFA pulse becomes apparent, modifying the transverse structure of the wake and leading to multiple transverse periods. In principle, both the temporal and transverse profile of the DLWFA pulse could be further engineered using select transverse modes and unequal mode weighting.

\section{Conclusions and Prospects}

A laser pulse composed of plasma waveguide modes with appropriately chosen frequencies can drive a wakefield at the vacuum speed of light, enabling electron energy gains beyond the dephasing limit. The acceleration length of this DLWFA scheme increases with the number of modes in the pulse and depends on the electron density. This contrasts with standard LWFA, where the length depends only on density. The scaling with the number of modes provides additional flexibility to optimize either for higher energy gains at fixed stage length or for identical energy gains in shorter stages. The energy required in the DLWFA pulse scales proportionally with the stage length, such that, in principle, the pulse does not incur an energy penalty relative to standard LWFA.

The parameters for the simulation examples were chosen to verify the scaling laws while keeping the computational cost manageable. As a result, despite the $11\times$ increased energy gain enabled by DLWFA, the absolute energy gain was 660 MeV. The potential impact of DLWFA for high-energy colliders or radiation sources can be illustrated by comparison with a 100-GeV standard LWFA design consisting of ten 10-GeV stages. For $a_0 = 0.5$ and $\lambda_0 = 1 \, \mu\mathrm{m}$, such a design requires a density $n_\mathrm{S}=3.5\times10^{16} \,\mathrm{cm}^{-3}$ and a stage length $L_\mathrm{S} = L_\mathrm{d} = 5.7 \, \mathrm{m}$ [Eq.~\eqref{eq:dgS}]. Using a ten-mode pulse, DLWFA could achieve 100 GeV in only five stages operating at   $n_\mathrm{D}=2\times10^{17} \,\mathrm{cm}^{-3}$, with a stage length $L_\mathrm{D} = 4.7 \, \mathrm{m}$ [Eqs.~\eqref{eq:dgD} and \eqref{eq:La}]. The higher operating density reduces the total pulse energy by a factor of 2.4, $5 U_\mathrm{D} = (10U_\mathrm{S})/2.4$, and may also mitigate beamstrahlung at the interaction point \cite{Schroeder2010}. These advantages come at the expense of a larger matched spot size: $w_\mathrm{D} = 690 \, \mu\mathrm{m}$ compared with $w_\mathrm{S} \simeq 160 \, \mu\mathrm{m}$. However, this could also be a benefit by improving tolerances and reducing sensitivity to misalignment.

In addition to overcoming dephasing, a primary motivation for DLWFA in a plasma waveguide is to retain the velocity control afforded by flying-focus pulses while avoiding the large plasma volumes required to suppress refraction. To prevent refraction from altering the trajectory of an ultrashort flying focus or reducing energy coupling, the plasma volume $V$ should approximately fill the full aperture $f_\#$ cone of the axiparabola over the acceleration length. This yields the scaling $V \propto \pi L^3/(12f_\#^2)$. In contrast, a plasma waveguide confines the interaction to a narrow cylinder with $V\propto \pi Nw^2L = \sigma (\omega_\mathrm{p0}/\omega_0)^3 L^3/\pi$ [see Eqs.~\eqref{eq:wm} and~\eqref{eq:La}], where the factor of $N$ ensures that the waveguide radius ($\propto \sqrt N w$) is large enough to confine the $N-1$ mode. Because $\omega_\mathrm{p0}/\omega_0 \ll 1$, the waveguide could substantially reduce the required volume and ionization-energy. 

Realizing the advantages of DLWFA in a plasma waveguide requires generating a highly structued laser pulse. As discussed in \cite{Palastro2025}, a pulse composed of multiple spatial modes with distinct frequencies can be assembled using recently developed techniques for spatiotemporal structuring \cite{Cruz-Delgado2022,Piccardo2023,Zhan24}. One approach is to disperse a broadband pulse into several frequency bands that are independently tailored using metasurface optics or a spatial light modulator \cite{Cruz-Delgado2022,Piccardo2023,Zhan24}. Alternatively, the multiplexed outputs of a high-power fiber laser \cite{Rainville2024} could be independently manipulated before coherent combination. Recent advances in plasma and gas-phase optics \cite{Edwards2022} suggest a third route, in which an appropriately designed diffractive element separates different frequency bands of a high-power pulse into distinct spatial modes.

A laser pulse with an intensity peak that travels at or near the vacuum speed of light in a plasma channel can also be used to overcome dephasing in other LWFA schemes, such as multi-pulse LWFA \cite{Hooker2014,Jakobsson2021}. Following the design procedure described in Sec. III, a configuration in which multiple luminal intensity peaks resonantly excite the wake would be guided by two considerations: (1) choosing the FWHM duration to nearly maximize the wake strength, $\tau = \pi\sigma/\omega_\mathrm{p0}$, and (2) setting the recurrence period to $T_\mathrm{R} = 2\pi\ell/\omega_\mathrm{p0}$, where $\ell$ is an integer. Together with Eqs.~\eqref{eq:TR} and~\eqref{eq:fwhm}, these conditions yield the simple relation $N = 2\ell$. The resulting pulse would combine resonant wakefield enhancement with dephasing-free acceleration. More generally, the pulses described here may prove useful in the wide range of plasma-based applications that benefit from the flying focus.

\appendix

\section{Spatiotemporal Profile}

The spatiotemporal profile of the laser pulse is obtained by substituting Eq.~\eqref{eq:omgq} into Eq.~\eqref{eq:FT} and Taylor expanding $k_{q}(\omega)$ about $\omega = \omega_{q}$ to first order. This yields
\begin{equation}\label{eq:STP}
\begin{aligned}
|a(\mathbf{x},t)|^2 &= \frac{1}{N^2} |\alpha(t-z/v_{\mathrm{g}})|^2 \\ \bigg| \sum_{q=0}^{N-1} &L_q(\tfrac{2r^2}{w^2})\mathrm{e}^{-r^2/w^2}  
\exp\left[-iq\Omega(t-z/c) \right] \bigg|^2,
\end{aligned}
\end{equation}
where $\alpha(t) = \tfrac{1}{2\pi}\int \tilde{\alpha}(\omega)\mathrm{e}^{-i\omega(t-t_\mathrm{d})}d\omega$ and the phases $\Phi_q = q\Omega t_\mathrm{d}$ have been chosen to allow for a desired delay $t_\mathrm{d}$. In the regime considered here [Eq.~\eqref{eq:cond}], the group velocities of the modes, $v_{\mathrm{g},q} = c^2k_q(\omega_q)/\omega_q$, are nearly equal, so that $v_{\mathrm{g},q} \approx v_{\mathrm{g},0} \equiv v_\mathrm{g}$. For simplicity, the sum is taken over consecutive modes, although this is not required. On axis ($r=0$), $L_q(\tfrac{2r^2}{w^2})\mathrm{e}^{-r^2/w^2} = 1$ and the summation reduces to a finite geometric series. Evaluating the series leads to Eq.~\eqref{eq:STPOA}. Note that, due to the non-zero frequency spread of $\tilde\alpha(\omega)$, the shape-invariant property described by Eq.~\eqref{eq:cond} only occurs within the uniform intensity region of the temporal envelope  $\alpha(z-v_\mathrm{g}t)$.

\section{Simulation Details}
The simulations presented in this work were performed using \textsc{osiris} \cite{Fonseca2002}. A moving window and quasi-3D geometry were employed, with the fields of the laser pulse and plasma decomposed into a truncated expansion of azimuthal modes \cite{Davidson2015}. A customized field solver that mitigates errors from numerical dispersion and the time staggering of the electromagnetic fields was used \cite{Li2021}. The domain of the moving window extended $178~\mu$m longitudinally and $323~\mu$m radially, corresponding to $4980 \times 1490$ cells in $z$ and $r$, respectively. The zeroth- and first-order azimuthal modes were retained, which was sufficient to capture the salient dynamics for the linearly polarized laser pulse. The time step was 70.9~as, and the total simulation time was 109~ps. The electrons were represented using 16 particles per cell. The longitudinal density profile of the plasma had an initial $152$-$\mu$m upramp followed by a uniform region. The temporal profile of the DLWFA pulse, $\alpha$, consisted of a 36-fs rise, a 258-fs flattop, and a 36-fs fall, resulting in an intensity FWHM of 286~fs. The rise and fall were sixth-order polynomials with vanishing derivatives at their end points.

\begin{acknowledgments}
The authors would like to thank Jessica Shaw, Jeremy Pigeon, and Dustin Froula for discussions. The work of J.P.P., K.G.M., C.D.A., A.L.E., A.K., L.S.M., and D.R. is supported by the Office of Fusion Energy Sciences under Award Numbers DE-SC0021057, the Department of Energy National Nuclear Security Administration under Award Number DE-NA0004144, the University of Rochester, and the New York State Energy Research and Development Authority. The work of A.G.R.T is supported by the U.S. National Science Foundation under Award Number 2512014. This report was prepared as an account of work sponsored by an agency of the US Government. Neither the US Government nor any agency thereof, nor any of their employees, makes any warranty, express or implied, or assumes any legal liability or responsibility for the accuracy, completeness, or usefulness of any information, apparatus, product, or process disclosed, or represents that its use would not infringe privately owned rights. Reference herein to any specific commercial product, process, or service by trade name, trademark, manufacturer, or otherwise does not necessarily constitute or imply its endorsement, recommendation, or favoring by the US Government or any agency thereof. The views and opinions of authors expressed herein do not necessarily state or reflect those of the US Government or any agency thereof.
\end{acknowledgments}

\bibliography{main} 

\end{document}